\documentclass[twocolumn,pra,aps]{revtex4}
\usepackage{graphicx}

\begin{document}

\title{Time and nonlocal realism:\\Consequences of the before-before experiment}

\author{Antoine Suarez}
\address{Center for Quantum Philosophy, P.O. Box 304, CH-8044 Zurich, Switzerland\\
suarez@leman.ch, www.quantumphil.org}

\date{August 15, 2007}

\begin{abstract}It is argued that recent experiments refuting nonlocal realism, can also be considered as experiments refuting \emph{time-ordered nonlocality} and, hence, confirming the result of the \emph{before-before} experiment. However, the \emph{before-before} experiment provides a broader refutation because it also falsifies the testable relativistic version of Bohm's nonlocal model. All this stresses the interest of a new before-before experiment demonstrating together the failure of time-ordered nonlocality and the violation of the Leggett's inequality.\\


\end{abstract}


\maketitle

\section* {Introduction}

The concept of ``nonlocal realism'' has been introduced to characterize theories that pretend to explain quantum entanglement assuming both nonlocal influences and realism \cite{grö,le}. Nonlocal influences means actions producing correlated events in two space-like separated regions. Realism is the viewpoint according to which the results of observations are a consequence of pre-existing properties carried by physical systems. One can define nonlocal realistic theories that fulfill the so called Legget's inequality, whereas quantum mechanics violates it \cite{le}. Gr\"{o}blacher \emph{et al.} have presented experimental results violating Leggett's inequality, and in agreement with quantum mechanics. They concluded that ``giving up the concept of locality is not sufficient to be consistent with quantum experiments, unless certain intuitive features of realism are abandoned'' \cite{grö}. Meanwhile, other more conclusive experiments falsifying Leggett's nonlocal hidden variables model have been done \cite{vs,az}.

Relativistic quantum experiments with beam-splitters in motion were proposed in 1997 \cite{asvs97, as00.1} with the aim of testing the assumption that entanglement can be explained by means of \emph{time-ordered nonlocal influences}. The tested nonlocal model keeps the relativity of time, and uses the inertial frames of the beam spitters to establish the time order of the events. A consequence of this assumption is that the quantum correlations should disappear in relativistic experiments in which each beam-splitter, in its own reference frame, is first to select the output of the photons (before-before timing). This prediction was proved wrong by the \emph{before-before experiment} performed in 2001 \cite{szsg}.
Reference \cite{vs} quotes this experiment as the first falsification of a nonlocal hidden variables model, the ``Suarez and Scarani model'' \cite{asvs97}.

Here I show that the concept of ``nonlocal realism'' is equivalent to the assumption of time-ordered nonlocal influences, and the experimental falsification of nonlocal realistic models is a confirmation of the before-before experiment. I argue moreover, that this experiment actually also falsifies the relativistic testable version of Bohm's nonlocal theory. All this stresses the interest of a future before-before experiment demonstrating together: the violation of the Bell's and Leggett's inequalities, and the non disappearance of the nonlocal correlations.

\section* {Nonlocal realistic theories}

The nonlocal realistic theories fulfilling the Leggett's inequality are characterized through the following description (see \cite{grö}, and the corresponding supplementary information):

A source emits pairs of photons in a maximally entangled state, and it is assumed that the single photons in the pair carry well-defined polarizations. One of the photons with polarization vector \textbf{u} is sent to Alice's laboratory and measured with a polarizing beam-splitter set at angle \textbf{a}, and the other photon with polarization vector \textbf{v} is sent to Bob's laboratory and measured with a polarizing beam-splitter set at angle \textbf{b}. A polarization measurement gives either a result of $+1$ or $-1$ depending on whether a single photon is transmitted or reflected by its polarizer.

For reasons of clarity, Gr\"{o}blacher \emph{et al.} choose an explicit non-local dependence of Bob's outcomes on Alice's ones, though, they note, that one can also choose any other example of a possible non-local dependence. Thus, the local polarization measurement outcomes A are predetermined by the polarization vectors \textbf{u} and an additional set of hidden variables $\lambda$ specific to the source. The local polarization measurement outcomes B are predetermined by the polarization vectors \textbf{u} and \textbf{v}, the set of hidden variables $\lambda$, the settings \textbf{a} and \textbf{b}, and any possible non-local dependence of Bob's outcomes on Alice's ones.

It is a crucial trait of the Leggett's nonlocal models that there exist subensembles of definite polarizations before measurements, described by a probability distribution $\rho_{\textbf{\footnotesize u}, \textbf{\footnotesize v}}(\lambda)$. Then, the non-signalling condition imposes that the local averages performed on the subensemble of definite (but arbitrary) polarizations \textbf{u} and \textbf{v} obey Malus' Law, i.e.:

\begin{equation}\label{A}
    \bar{A}(\textbf{u})=\int  A(\textbf{a},\textbf{u},\lambda) \rho_{\textbf{\footnotesize u}, \textbf{\footnotesize v}} (\lambda) d\lambda = \textbf{u} \cdot \textbf{a}
\end{equation}

\begin{equation}\label{B}
    \bar{B}(\textbf{v})=\int  B(\textbf{a},\textbf{b},\textbf{u},\textbf{v},\lambda) \rho_{\textbf{\footnotesize u}, \textbf{\footnotesize v}} (\lambda) d\lambda = \textbf{v} \cdot \textbf{b}
\end{equation}

All nonlocal dependencies are put on the side of Bob, in Equation (\ref{B}): his measuring device has the information about the setting of Alice, \textbf{a}, and her polarization \textbf{u}. Theories in accordance with this description fulfill the Leggett's inequality \cite{grö}, while they violate the Bell's inequality to the same extent of quantum mechanics. Both inequalities are tested by performing coincidence measurements (A,B) with different settings. The experiments show violations of Leggett's and Bell's inequalities, in agreement with the predictions of quantum mechanics \cite{vs,az}.

\begin{figure}[t]
\includegraphics[width=80 mm]{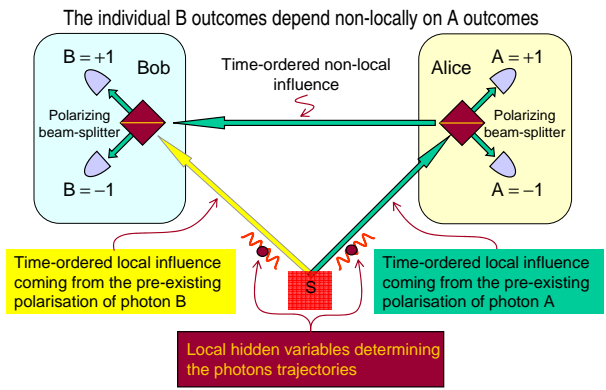}
\caption{The Legget's inequality is a characteristic of causal chains built of time-ordered nonlocal influences and time-ordered local ones. The figure shows (blue-green, counterclockwise) the causal chain corresponding to the choice of Reference \cite{grö}: the individual B outcomes depend nonlocally on pre-existing A outcomes, and these depend locally on the polarization of photon A when it leaves the source; a similar (not shown) causal chain could be drawn (yellow, clockwise) for the alternative case of the individual A outcomes depending nonlocally of B outcomes.}
\label{fig1}
\end{figure}

\section* {The core of nonlocal realism consists in time-ordered nonlocal influences}

Let us now consider the postulate of ``nonlocal realism'' \cite{grö,le}. According to this postulate all measurements A and B are determined by the ``pre-existing'' properties (hidden variables) the particles carry. However, the individual B outcomes cannot be considered to be predetermined by the polarization vectors \textbf{v} the photons B carry when they leave the source and the set of local hidden variables $\lambda$ specific to the source, for the B outcomes depend non-locally on the A outcomes, as Equation (\ref{B}) indicates. The influence coming from the pre-existing polarization of photon B does not matter for the value of the individual B outcomes, but only for their statistical distribution. Therefore, ``non-local realism'' implies that the individual outcomes in Bob's laboratory are determined by ``pre-existing'' space-like separated outcomes in Alice's laboratory, and these are determined by the hidden variables the corresponding photons A carry when they leave the source, according to the causal chain sketched in Figure 1. Other features like the trajectories the photons travel from the source to the polarizing beam-splitters, are exclusively determined  by local realistic properties. But even the Copenhagen interpretation of Quantum Mechanics admits some local realistic features like, for instance, the property that a detection takes place \emph{D/c} seconds after the time of emission, where \emph{D} means the distance between source and the detectors, and \emph{c} the velocity of light.

The term ``pre-existing'' has an obvious \emph{temporal} meaning. Hence, for coincidence measurement outcomes (A, B) in the laboratories of Alice and Bob, ``nonlocal realism'' means that each individual outcome A at Alice's side precedes in time and causes the correlated individual outcome B at Bob's side. Thus, one is led to the conclusion that the postulate of ``nonlocal realism'' requires the assumption of \emph{time-ordered nonlocal influences}. ``Nonlocal realism'' (as defined in \cite{grö, le}) fails if experiment proves wrong that \emph{one of two non-locally correlated events occurs before and is the cause of the other}.

\section* {The falsification of Leggett's nonlocal model is a confirmation of the before-before experiment}

Time-ordered nonlocal influences imply that it is possible to establish the timing of the outcomes by means of a \emph{real} clock. The trouble is that since Bob and Alice stay space-like separated the timing will depend on the inertial frame of the measuring observer. However, any ``realistic'' theory must assume that the photon always travel a definite trajectory determined by the particle's properties, and the outcome becomes determined at the corresponding polarizing beam-splitter at the moment the particle hits it (Figure 1); additionally, a photon experiencing the inertial frame of the beam-splitter becomes doppler shifted when it is reflected by a beam-splitter in motion. Therefore it is logical to assume that the inertial frame of Alice's polarizing beam-splitter defines the clock measuring the time for her outcome, and the inertial frame of Bob's polarizing beam-splitter defines the clock measuring the time for his outcome.

Consider now a relativistic experiment with beam-splitters in motion in such a way that each of them, in its own reference frame, is first to select the output of the photons (before-before timing). Then, each outcome will become independent of the other, and the nonlocal correlations should disappear. This means that theories assuming time-ordered non-local influences predict the disappearance of non-local correlations in before-before experiments, and the same holds for theories sharing ``non-local realism''. A before-before experiment has been done in 2001 \cite{szsg}, though not using polarizers but interferometers. The result was that the correlations doesn't disappear. Hence, the experiment ruled out time-ordered nonlocal influences, and thereby proved ``nonlocal realism'' wrong. None of the two correlated events can be considered the cause of the other. In words of Nicolas Gisin: ``the same randomness manifests itself at several locations'' \cite{ng}. A quantum correlation is a single event in which local randomness and non-local order (correlations) appear inseparably united. Entanglement has its roots outside of space-time \cite{as07}. The joint outcomes in entanglement experiments cannot be described as being determined by pre-existing properties of particles independent of the measurement.

Conversely, the experimental falsification of Leggett's nonlocal model means the falsification of time-ordered nonlocality. Although the inequality (as said above) characterizes theories combining time-ordered nonlocal influences together with certain features of local realism, these realistic features are supposed to hold in certain cases and fail in others: so for instance, the individual A outcomes (Figure 1) can be considered completely determined by pre-existing properties photon A carries, but the individual B outcomes are not completely determined by pre-existing properties photon B caries. In this sense ``the abandon of certain intuitive features of realism'' enters already as an axiom the non-local hidden variables theories, rather than being a consequence of the experiment. This means that the core of Leggett's nonlocal model is time. Time-ordered nonlocal influences are a sort of ``hidden'' assumption in the nonlocal hidden variables models. The effective consequence of the violation of the Leggett's inequality is the failure of a particular type of time-ordered nonlocality.

In any case the violation of Legget's inequality does not imply at all that a conscious human observer has to be present for the collapse of the wave function to take place.

\section* {The violation of Leggett's inequality does not rule out Bohm's nonlocal model}

Alain Aspect commenting on the experiment of Gr\"{o}blacher \emph{et al.} stresses: ``there are other types of non-local models that are not addressed by either Leggett's inequalities or the experiment.'' \cite{aa} Aspect considers the following model: Assume that a measurement is performed first on photon A. This measurement gives either a result of $+1$ or $-1$; immediately after one of the two results is obtained, the quantum description of photon B, which had not been favoring any precise polarization before the measurement on photon A, collapses into a state of polarization identical to the one found for photon A, from which one can readily derive the usual quantum-mechanical EPR correlations. As Aspect states, this model is clearly nonlocal in the relativistic sense, as we must invoke a particular frame of reference to give a sense to the statements that measurement on photon A happens first, and that its result immediately affects the state of photon B. In a sense the model is also realist, as we can qualify the individual polarization of each photon at each step \cite{aa}. Aspect claims: ``If we take this description - based on standard quantum-mechanical calculations - as a model, it cannot be rejected by any experiment that is in agreement with quantum mechanics, including the more complex elliptical polarization measurements performed by Gr\"{o}blacher \emph{et al}.'' \cite{aa}. Aspect's explanation is actually nothing other than that provided by Bohm's theory \cite{db}, and also the way John Bell used to describe non-locality \cite{jb87}. It could be sketched deleting in Figure 1 the two arrows corresponding to the local influences coming from pre-existing photons polarizations.

Gr\"{o}blacher \emph{et al.} themselves also acknowledge that their experiment does not rule out Bohm's model. Indeed, in this model neither of the two photons in a maximally entangled states carries any definite polarization when leaving the source: they acquire it only when they pass the polarizing beam-splitters through the action of the nonlocal ``quantum potential'' \cite{dhk}. Fulfilling Leggett's inequality is not a characteristic of Bohm's model \cite{grö,vs}.

Bohm's theory casts nonlocality into a time-ordered causal scheme. It postulates that the outcome at one beam-splitter precedes in time, and causes faster than light the outcome at the other beam splitter by means of the ``quantum potential''. However, Bohm gives up the relativity of time and postulates an \emph{undefined} preferred frame or universal clock to establish the time order.

Note that Gr\"{o}blacher \emph{et al.} claim that Bohm's theory is ``realistic'' despite its denial of ``pre-existing'' definite photon polarizations. This claim stresses that in the ``experimental tests of nonlocal realism'' the ``hidden'' hypothesis one wants to test is actually time-ordered nonlocality.

\section* {After the before-before experiment Bohm's interpretation can hardly be considered a valid alternative to Copenhagen}

Bohm's theory is an attempt to give a complete deterministic interpretation of quantum mechanics.

However, the free will of the experimenter is a main assumption in Bell's proof of nonlocality \cite{as07}: this feature remains often hidden in the discussion of the experiments. Every nonlocal theory is based on the axiom that important parts of the physical world, the human brains, cannot be exclusively explained in a deterministic way, and free choices are possible. The mere assumption of nonlocal influences throws out determinism. Hence, it is inconsistent to claim that Bohm's  model is both nonlocal and fully deterministic.

Can we consider Bohm's model at least \emph{nonlocal deterministic} in the sense of time-ordered nonlocal causality? \cite{as07}. Bohm's time-ordered nonlocal theory is actually not testable, because it does not state which clock has to be used for establishing the time order of the events. For a scientific theory this is a severe weakness. If one assumes that Bohm's time-ordered nonlocality is not an illusion but belongs to physical reality, one has to cast it into a description using \emph{real} clocks. As said above, the essential ingredients of a realistic theory lead naturally to accept that the relevant clocks are those defined by the inertial frames of the beam splitters. In this sense, Bohm's model is the adequate time-ordered nonlocal description for entanglement experiments with beam splitters at rest. And its natural extension to entanglement experiments with beam splitters in motion is the \emph{multisimultaneity} (``Suarez and Scarani'') model, leading to the prediction that the correlations should disappear in the before-before experiment \cite{szsg, as07}. Therefore, this experiment proves nonlocal determinism in the testable relativistic extension of Bohm's model wrong.

Nonlocality assumes free choices on the part of the experimenter, and the before-before experiment demonstrates nonlocal free choices on the part of Nature: Bohm's deterministic interpretation can hardly be considered a valid alternative to the non-deterministic view of Copenhagen \cite{as07}.

Therefore the before-before experiment means a broader refutation of nonlocal determinism (time-ordered nonlocality) than that provided by the experimental violation of Legett's inequality. Nevertheless, it is noteworthy that this inequality offers a mean of testing a particular case of time-ordered nonlocality without the necessity of setting devices in motion to achieve relativistic timing configurations.

\section* {Conclusion and proposal}

The preceding analysis shows that what one tests with Leggett's inequality is not realism but time, time-ordered nonlocality. The before-before experiment ruled out theories assuming time-ordered nonlocal influences. The experimental tests of Leggett's nonlocal models versus quantum mechanics confirm this result. Giving up the concepts of locality and realism is not sufficient to be consistent with quantum experiments, one has also to abandon time-ordered causality too.

This issue has not yet been highlighted in the ongoing discussion about experimental tests of nonlocal realistic theories. This may be a sign that ``most working scientists'' (the expression appears in \cite{grö}) unconsciously still hold fast to the postulate of time-ordered causality: Apparently, the idea that Nature establishes order without time is, after all, the most counterintuitive feature of quantum mechanics. A possible way of rendering this idea more familiar, and contributing to a better understanding of the nature of quantum entanglement, could be demonstrating within the same experiment all the three results: the violation of Bell's and Leggett's inequalities, and the non-disappearance of the quantum correlations with before-before timing.

\emph{Acknowledgments}: I am grateful to Nicolas Gisin, Valerio Scarani, Andr\'{e} Stefanov and Hugo Zbinden for inspiring discussions.


\begin{references}


\bibitem{grö}  Gr\"{o}blacher S.,  Paterek T.,  Kaltenbaek R., Brukner \v{C}., \.{Z}ukowski M., Aspelmeyer M.,  Zeilinger A., An experimental test of non-local realism. \emph{Nature} \textbf{446}, 871-875 (2007) .

\bibitem{le} Leggett A. J., Nonlocal hidden-variable theories and quantum mechanics: An incompatibility theorem. \emph{Found. Phys.} \textbf{33}, 1469–1493 (2003).

\bibitem{vs} Branciard C.,  Ling A.,  Gisin N., Kurtsiefer Ch., Lamas-Linares A.,  Scarani V., Experimental Falsification of Leggett's Non-Local Variable Model. 	 \emph{arXiv:0708.0584v1 [quant-ph]} (2007).

 \bibitem{az} Paterek T., Fedrizzi A., Gr\"{o}blacher S.,    Jennewein Th., \.{Z}ukowski M., Aspelmeyer M.,  Zeilinger A., Experimental test of non-local realistic theories without the rotational symmetry assumption.	 \emph{arXiv:0708.0813v1 [quant-ph]} (2007).

\bibitem{asvs97} Suarez A. and Scarani V., Does entanglement depend on the timing of the impacts at the beam-splitters? {\em Phys. Lett. A}, {\bf 232}, 9 (1997).

 \bibitem{as00.1} Suarez A., Quantum mechanics versus multisimultaneity in experiments with acousto-optic choice-devices. {\em Phys. Lett. A}, {\bf 269}, 293 (2000).

\bibitem{szsg}  Stefanov A.,  Zbinden H.,  Gisin N., and  Suarez A., Quantum Correlations with Spacelike Separated Beam Splitters in Motion: Experimental Test of Multisimultaneity. {\em Phys. Rev. Lett.} \textbf{88} 120404 (2002). Quantum entanglement with acousto-optic modulators: 2-photon beatings and Bell experiments with moving beamsplitters \emph{Phys. Rev. A} \textbf{67}, 042115 (2003).

\bibitem{ng} Gisin N. Can relativity be considered complete? From Newtonian nonlocality to quantum nonlocality and beyond. \emph{quant-ph/0512168v1} (2005).

\bibitem{as07} Suarez A., Entanglement and time.  \emph{quant-ph/0311004} (2003). Classical demons and quantum angels. \emph{arXiv:0705.3974v1 [quant-ph]} (2007).

\bibitem{db} Bohm D., A suggested interpretation of the quantum theory in terms of ``hidden'' variables. I and II. \emph{Phys. Rev.} \textbf{85} 166-193 (1952).

\bibitem{jb87} Bell J.S., {\em Speakable and unspeakable in quantum mechanics}, Cambridge: University Press, 1987.

\bibitem{dhk} Dewdney C., Holland P.R., Kyprianidis A., Vigier J.P., Spin and non-locality in quantum mechanics. \emph{Nature} \textbf{336}, 536–544 (1988).

\bibitem{aa} Aspect A., To be or not to be local. \emph{Nature}, \textbf{446}, (2007) 866-867.


\end{references}
\end{document}